\documentclass[useAMS,usenatbib]{mn2e}
\usepackage{epsfig}
\title{Kinematics of Milky Way Satellites in a Lambda Cold Dark Matter Universe}

\author[Strigari et al.] {\parbox{18cm}{
Louis~E.~Strigari$^{1}$, Carlos~S.~Frenk$^{2}$ and Simon~D.~M.~White$^{3}$ 
}\vspace{0.3cm}\\
$^1${Kavli Institute for Particle Astrophysics and Cosmology, Stanford University, Stanford, CA 94305 USA}\\
$^3${Institute for Computational Cosmology, Dep. of Physics,
    Univ. of Durham, South Road, Durham  DH1 3LE, UK}\\
$^{3}${Max-Planck-Institut f\"{u}r Astrophysik,
Karl-Schwarzschild-Stra\ss{}e 1, 85740 Garching bei M\"{u}nchen,
Germany}\\
}

\begin{document}

\maketitle

\begin{abstract}
We investigate whether the subhalos of $\Lambda$CDM galaxy halos have
potentials consistent with the observed properties of Milky Way
satellites, particularly those with high-quality photometric and
kinematic data: Fornax, Leo I, Sculptor, Sextans, and Carina.  We
compare spherical models with isotropic velocity dispersion tensors to
the observed, circularly averaged star counts, line-of-sight velocity
dispersion profiles and line-of-sight velocity distributions.  We
identify subhalos within the six high-resolution dark matter halos of
the Aquarius Project for which the spherically averaged potentials
result in excellent fits to each of the five galaxies.  In particular,
our simple one-integral models reproduce the observations in the inner
regions, proving that these data are fully consistent with
$\Lambda$CDM expectations and do not require cored dark matter
distributions. For four of the five satellites the fits require
moderately cusped {\it stellar} density profiles. The star count data
for Leo~I, however, do require a cored distribution of star counts.
Current data suggest that these five satellites may be hosted by
$\Lambda$CDM subhalos with maximum circular velocities in the range 10
to 30~km~s$^{-1}$.
\end{abstract} 


\section{Introduction}

The internal dynamics of the dwarf spheroidal (dSph) satellite
galaxies of the Milky Way (MW) offer perhaps the best prospects for
investigating the properties of the dark matter in the nearby
universe. These galaxies are dominated by dark matter and the
brightest of them are sufficiently close that high-precision
line-of-sight velocities for large samples of their stars can be
measured using high-resolution, multi-object spectroscopy
~\citep{Walker:2008fc}. The current datasets represent an improvement
upon the earliest such observations ~\citep{Aaronson1983,Mateo1993} by
factors of tens to thousands. Analysis of the kinematic data, in
combination with improved photometric measurements, have not only
confirmed earlier indications that the classical dSphs are dark matter
dominated, but have also revealed the surprising property that their
mean central densities are similar even though their luminosities span
a very wide range of
values~\citep{Mateo:1998wg,Walker2007,Gilmore2007,Strigari:2008ib}.

Current cosmogonic theory makes strong predictions for the internal
structure of dark matter halos. N-body simulations of halo formation
in hierarchical clustering cosmogonies have shown that halos develop
strongly cusped density profiles which are almost independent of halo
mass and cosmological parameters \citep{nfw96,nfw97}. Subsequent
simulations have confirmed this
result \citep[e.g.][]{Navarro:2008kc,Stadel:2009}, showing, in addition,
that cuspy profiles are retained even after halos fall into larger
ones and suffer extensive tidal stripping
\citep{Kazantzidis:2005su,springel2008}. Thus, to the extent that the
dark matter distributions in the inner parts of halos have not been
significantly disturbed by the galaxies forming within them, halo
profiles offer a strong and direct test of the $\Lambda$CDM cosmogony
in a regime not probed by microwave background and large-scale
structure data.

An in-depth analysis of the central regions of Milky Way
satellites is particularly important given that a number of recent
studies of the structure of these galaxies have claimed that shallow
central density profiles provide a better description of their dark
matter halos than the cuspy profiles characteristic of $\Lambda$CDM
\citep{Goerdt:2006rw,sanchez2006,Gilmore2007}. Models with dark
matter core radii of $\sim 100$~pc have been shown to provide good
fits to the kinematic data sets of the classical dSphs
\citep{Angus:2009jh}. If confirmed, the shallow cores suggested by
these studies might indicate a lower central phase-space density than
expected if the dark matter is a cold collisionless particle
\citep{Tremaine:1979we,Hogan:2000bv}.

Kinematic studies typically treat dSph galaxies as spherical,
dynamically equilibrated systems. With these simple assumptions there
is a strong degeneracy between the statistics of stellar orbits
(i.e. whether velocity dispersions are isotropic, or are radially or
tangentially biased) and the shape of the stellar and dark matter
density profiles \citep[e.g.][]{Evans:2008ik}. This ambiguity is
reflected in the broad range of models used in recent attempts to
constrain the dark matter density profiles of the dSphs
\citep{Strigari:2008ib,Lokas:2009cp,Walker:2009zp,Wolf:2009tu}. The
intrinsic parameter degeneracies of the models cast doubt on the
robustness of inferences favoring cored or cuspy central density
profiles, even given the high-quality data that are now
available \citep{Strigari:2006ue,Walker:2009zp}. Breaking these
degeneracies may only be possible by exploiting additional
observational constraints, for example, through measurement of
internal stellar proper
motions \citep{Wilkinson:2001ut,Strigari:2007vn}.

Motivated by the simple question of whether the observed dSphs are
kinematically consistent with the $\Lambda$CDM theory of structure
formation, we here use six high-resolution halo simulations performed
as part of the Aquarius Project \citep{springel2008} to search for
subhalos whose properties would allow them, in principle, to host the
well observed dSph satellites of the Milky Way. We restrict our
attention to the five satellites with abundant, high-quality stellar
kinematic data: Fornax, Sculptor, Leo I, Carina, and
Sextans \citep{Mateo:2007xh,Walker:2008fc}.  Assuming isotropic
velocity dispersions, we identify an Aquarius subhalo with a dark
matter potential which results in a good simultaneous fit to each
satellite's photometric and kinematic data.

In addition to focusing specifically on $\Lambda$CDM and making direct
use of realistic potentials from the Aquarius simulations, our
modeling differs from previous work in that we simultaneously fit both
photometry and kinematics. We allow for mildly cusped {\em stellar}
profiles with $\rho_\star \sim r^{-a}$ near the centre, where
$a$ is in the range 0 to 1. Projections of such profiles are
fitted to the photometric data, and with the results in hand we make
predictions for the kinematic data, both traditional second moment
(line-of-sight velocity dispersion) profiles and full line-of-sight
velocity distributions. The latter comparison allows us to test
whether simple, spherical, isotropic, $\Lambda$CDM-based models are
consistent with higher moments of the observed velocity distribution.

For each of the five satellites we study, the best-fitting Aquarius
subhalo provides an excellent statistical fit to the data. For four of
the five this requires a mildly cusped stellar distribution similar to
those found in brighter early-type galaxies. The only galaxy that
requires a true core in the star distribution is Leo~I, but with such
a profile its kinematics are still consistent with a $\Lambda$CDM
subhalo.  We present circular velocity curves for the best-fitting
subhalo hosts for each of the MW satellites. Not surprisingly, for a
given satellite, the circular velocity curves of ``good'' subhalos are
very similar at the radii that are well sampled by the stellar
tracers. This is a consequence of our assumptions of spherical
symmetry and isotropy.  Finally, we determine the mass of the
best-fitting subhalos, both at the time of accretion onto the host
halo and at high redshift, and we show that these quantities have more
scatter than the present-day central potentials or maximum circular
velocities.

\section{Theoretical Modeling} 
\label{sec:modeling} 
The goal of our theoretical modeling is ultimately to compare to the
full observed line-of-sight velocity distribution for each satellite.
We begin by discussing how to model velocity dispersion profiles and
then move on to modeling of the full line-of-sight velocity
distribution. In our later analysis, we will use predictions for
line-of-sight velocity dispersion profiles to identify possible
subhalo hosts for each satellite, and then use the corresponding
velocity distributions to check that our isotropic models are indeed
consistent with the observed kinematics.

\subsection{Velocity dispersion} 
We assume spherical symmetry both for the potential and for the
distribution of stars within it. The line-of-sight velocity dispersion
of a dSph satellite at projected radius, $R$, can then be written as
\begin{equation}
\sigma_{los}^2(R) = \frac{2}{I_\star(R)}
\int_R^\infty \left[1-\beta(r)\frac{R^2}{r^2}\right]
\frac{\rho_\star(r) \sigma_r^2 ~r}{\sqrt{r^2-R^2}} dr,
\label{eq:jeans}
\end{equation}
where $\rho_\star(r)$ is the stellar density profile and $I_\star(R)$
its two-dimensional projection; $\beta=1-\sigma_t^2/\sigma_r^2$ is the
anisotropy parameter, where $\sigma_t^2(r)$ is the one-dimensional
tangential velocity dispersion of the stars and $\sigma_r^2(r)$ the
corresponding radial velocity dispersion. These quantities satisfy the
radial Jeans equation:
\begin{equation} 
r \frac{d(\rho_{\star} \sigma_r^2)}{dr} =  - \rho_{\star}(r)GM(r)/r- 2
\beta(r) \rho_{\star} \sigma_r^2. 
\label{eq:jeansDE}
\end{equation}   
The total mass distribution, $M(r)$, is the sum of the mass
distributions of dark matter, $M_{dm}(r)$, and stars, $M_\star(r)$.

For our analysis, we use the dark matter mass distributions, $M_{dm}(r)$,
of subhalos in the next-to-highest resolution set of Aquarius
simulations of galactic halos (level 2 in the notation of
\citet{springel2008}). At the present day, each of the six halos
contains about 45000 resolved subhalos within $\sim 400$~kpc of its
centre (corresponding to the radius, $r_{50}$, of the sphere of mean
overdensity 50 times the critical value) with more than 20 particles,
each of mass $\sim 10^4 M_\odot$. The structure of each subhalo is
well resolved down to a physical radius of $\sim 100$ pc,
corresponding approximately to twice the gravitational softening
length~\citep{springel2008}. For radii greater than this convergence
radius and less than the radius where the density of bound subhalo mass
drops to $\sim 80\%$ of the local total mass density,
the subhalos are well fit by an ``Einasto'' profile,
\begin{equation}
\ln [\rho(r)/\rho_{-2}] = (-2/\alpha)[(r/r_{-2})^\alpha-1]. 
\label{eq:einasto}
\end{equation}
Here $\rho_{-2}$ and $r_{-2}$ are the scale density and scale radius
(at the point where the density profile has the isothermal slope) respectively,
and we take $\alpha=0.17$~\citep{Navarro:2008kc}. In our kinematic
analysis, we extrapolate an Einasto fit to each subhalo when it
is necessary to evaluate the mass distribution at radii
$< 100$ pc, and we use the directly determined mass profiles at all
larger radii.

To model the three-dimensional stellar density profile, $\rho_\star$,
we use functions of the form \citep{Zhao:1996}:
\begin{equation} 
\rho_\star(r) \propto \frac{1}{x^a(1+x^b)^{(c-a)/b}} 
\label{eq:3Dzhao}
\end{equation} 
where $x=r/r_0$ and $\{a,b,c,r_0\}$ are free parameters that will be
estimated in the next section by fitting the observed surface density
profile of each satellite.  We focus on cuspy central profiles ($0\leq
a \leq 1$) because we find that they are required to fit the observed,
nearly flat velocity dispersion profiles if we assume isotropic
stellar velocity dispersions ($\beta(r) = 0$) and an Einasto halo
profile. Although such cuspy profiles have not been used previously in
studies of the MW satellites, they are, in fact, required to fit the
inner surface brightness profiles of elliptical galaxies of all
luminosities, including faint ones \citep{Gebhardt:1996} and so seem
{\it a priori} quite plausible for dSph galaxies also.

If the stellar mass is everywhere negligible compared to the dark
matter mass, the projected velocity dispersion in Eq.~\ref{eq:jeansDE}
is independent of the constant of proportionality in
Eq.~\ref{eq:3Dzhao} that sets the stellar mass-to-light ratio,
$M_\star/L_\star$. However, if the stars contribute significantly to
the potential of the galaxy, then we must determine the appropriate
normalizing factor for Eq.~\ref{eq:3Dzhao} and thus its contribution
to the overall mass distribution of the galaxy. For each of the dSphs
we will take $M_\star/L_\star = 1$, consistent with the observational
results~\citep{Mateo:1998wg,Coleman2005}.
We find that small variations in
$M_\star/L_\star$, indicative, perhaps, of multiple stellar
populations or differences in stellar initial mass function, have
little effect on the results we present below.

As also noted above, throughout our analysis we will assume locally
isotropic velocity distributions, $\beta(r)=0$ for all $r$. This is a
strong assumption and it is thus remarkable that we find that we can
fit all the kinematic data without relaxing it.

\subsection{Velocity Distributions} 
The preceding discussion demonstrates the well-known fact that the
observable quantities $I_*(R)$ and $\sigma_{los}^2(R)$ are
insufficient to determine the mass profile, $M(r)$, of a spherical
system unless the velocity anisotropy, $\beta(r)$, is specified.
Additional kinematic information is contained in higher order moments
of the line-of-sight velocity distribution, so appropriate modeling of
these moments may constrain $\beta(r)$ and so
$M(r)$~\citep[e.g.][]{Gerhard1993,Lokas:2003ks}.  A fully consistent
dynamical model must clearly match the full line-of-sight velocity
distribution at all radii.

If we assume $\beta(r)=0$, it is possible to invert the observables
$I_*(R)$ and $\sigma_{los}^2(R)$ to obtain not only a unique $M(r)$
but also the unique distribution function, $f(\epsilon)$, which
reproduces these observables within the potential corresponding to
$M(r)$. This distribution function then determines the full
line-of-sight velocity distribution at each $R$. Thus, once we have
found a subhalo with $M(r)$ consistent with the $I_*(R)$ and
$\sigma_{los}^2(R)$ measurements for a particular dSph, we can
check the consistency of the resulting model by comparing its
line-of-sight velocity distributions with those observed.

To obtain these velocity distributions, we begin with the Eddington
inversion formula,
\begin{equation} 
f(\epsilon) = \frac{1}{\sqrt{8}\pi^2} \int_\epsilon^0 
\frac{d^2 \rho_\star}{d\Psi^2}\frac{d\Psi}{\sqrt{\Psi -\epsilon}}, 
\label{eq:eddington}
\end{equation} 
where $\epsilon = \Psi(r) + v^2/2$ is the binding energy, $\Psi$ is
the gravitational potential, and $v$ is the modulus of the velocity.
Potentials for the stars and the dark matter can be separately
constructed numerically via the Poisson equation, $\nabla^2
\Psi_\imath = 4\pi G \rho_\imath$.  The indices on potential and
density represent a specific component, the dark matter or the stars.
The total potential is then the sum of the two.

The Eddington formula in Eq.~\ref{eq:eddington} determines the
velocity distribution as a function of the binding energy. However, to
compare to observations we need line-of-sight velocity distributions
for a set of circular annuli.  Defining $v_{los}$ as the component of
velocity along the line-of-sight and performing the appropriate
weighting over three-dimensional radii $r$, the distribution of
line-of-sight velocities at projected radius $R$ is given by
\begin{equation} 
\hat f(v_{los};R) \propto \int_R^{r_{los}} \frac{r dr}{\sqrt{r^2-R^2}} 
\int_{\Psi(r)+ v_{los}^2/2}^0 f(\epsilon) d\epsilon, 
\label{eq:fhat}
\end{equation}
where $r_{los}$ is defined by $2\Psi(r_{los}) = v_{los}^2$; for a
given velocity $v_{los}$, we determine $r_{los}$ via a
numerical root-finding algorithm.  The normalization of
Eq.~\ref{eq:fhat} will not be important for the purposes of our
discussion.

Once we have determined the velocity distribution in Eq.~\ref{eq:fhat}
it is straightforward to construct higher order moments of this
distribution. In particular, the $n^{th}$ moment of the distribution is
given by
\begin{equation}
\langle v_{los}^n; R \rangle = \frac{\int v_{los}^n \hat f(v_{los}; R) dv_{los} }
{\int \hat f(v_{los}; R) dv_{los}}, 
\label{eq:moment}
\end{equation} 
where, as in Eq.~\ref{eq:fhat}, we have explicitly written $\hat f$ as a
function of the line-of-sight velocity. As an example that will be
important for us below, the RMS velocity determined from
Eq.~\ref{eq:fhat} is $\sqrt{\langle v_{los}^2; R \rangle}$. We are
thus able to check our numerical calculation of the velocity
distribution function by comparing the RMS velocity determined from
Eq.~\ref{eq:moment} to the equivalent quantity determined from
Eq.~\ref{eq:jeans}.

Eq.~\ref{eq:fhat} gives the theoretical velocity distribution
at $R$, but in practice, to compare to the observations, we must determine
the distribution of $\hat f$ at the position of each observed star,
and then average it over all the stars in each annulus.  Thus, our mean
$\hat{f}(v_{los})$ for a given annulus is the mean of the values at the
position of the stars, with the individual $\hat f$ distributions all
normalized to unity.  With this procedure there are no approximations
related to finite bin size when comparing our theoretical model to the
data.

\section{Data Analysis} 
\label{sec:analysis} 
In this section we perform our analysis of the photometric and
kinematic data. We first discuss how we use the star
count data to identify parameter values for Eq.~\ref{eq:3Dzhao} that
describe each satellite well. We then describe our handling and
interpretation of the line-of-sight velocity data, and the way in
which we use these data to identify specific Aquarius subhalos that 
could host each satellite. In particular, we describe the criterion by
which we judge goodness-of-fit for a given satellite-subhalo match.

\subsection{Photometry} 
\label{sec:photometry}
We use photometric data from the following sources for the surface
density profiles of our five galaxies: Fornax \citep{Coleman2005};
Sculptor \citep{Battaglia:2008jz}; Sextans \citep{Irwin:1995tb}; Leo
I \citep{Smolcic:2007hk}; Carina \citep{Munoz:2006hx}. Traditional
fits to these data sets have used King or Plummer models, the latter
corresponding to $\{a,b,c\} = \{0,2,5\}$ in Eq.~\ref{eq:3Dzhao}. Such
fits typically fail to reproduce the measured star counts in the outer
regions~\citep{Irwin:1995tb}. In addition, they enforce a constant
density core which is consistent with star counts in some globular
clusters but not with photometry of the inner regions of brighter
early type galaxies, almost all of which show inner cusps
corresponding to $a$ values significantly greater than
zero~\citep{Gebhardt:1996}.

We take $\{a,b,c,r_0\}$ in Eq.~\ref{eq:3Dzhao} as free parameters to
be adjusted when fitting the observed star count profiles. We perform
a standard Abel projection of $\rho_\star(r)$ to obtain $I_*(R)$ and
we determine the free parameters for each satellite via a standard
$\chi^2$ minimization procedure. In performing these fits, we find
that there is a complex degeneracy in the space spanned by the four
parameters. Motivated by reasons that we discuss in detail in
Section~\ref{sec:results}, we focus on models in which the 3D stellar
profiles are characterized by a central shallow cusp and a relatively
sharp turnover to a steep outer power law.  For Sculptor, Carina,
and Sextans we will specifically adopt a central cusp with $a=0.5$,
and $b = 3$, while for Fornax, we will use $a=1$ and $b=4$. In Leo I,
the star counts do appear to require a core, and we adopt $a=0$ and a
similar transition to the steep outer power law.  With
$a$ and $b$ fixed {\it a priori}, we vary the remaining parameters,
$\{c,r_0\}$, in order to minimize $\chi^2$. In all cases this results
in reduced $\chi^2$ values near unity, indicating an acceptable fit,
and also within the 90\% c.l. of
the minimum values attainable by varying all four
parameters independently. The resulting parameter sets
are given together with the corresponding $\chi^2/(N_{bin} - 5)$ in
Table~\ref{tab:attributes}. Note that since our goal is to demonstrate
that the observations are consistent with simple spherical, isotropic
models within $\Lambda$CDM subhalos, it is not necessary for us to
choose the best-fit profile parameters; rather we need only show that
the parameters we do choose are consistent with the star count data.

\begin{figure}
\begin{center}
\begin{tabular}{c}
{\resizebox{6.5cm}{!}{\includegraphics{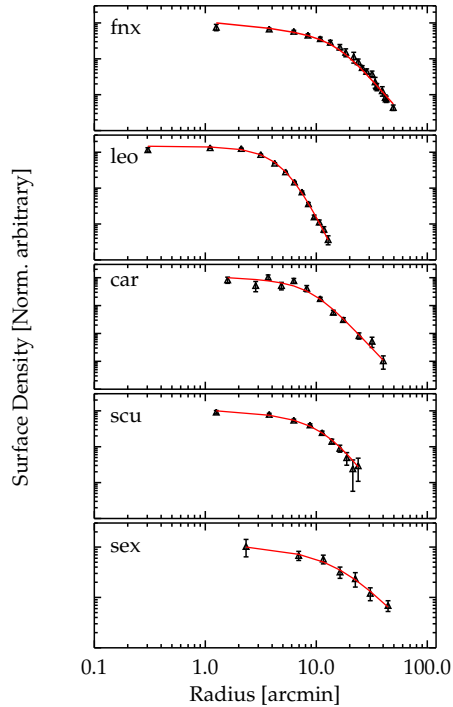}}}
\end{tabular}
\end{center}
\caption{Projected surface density profiles for each of the satellites
  that we consider, fit using the formula in Eq.~\ref{eq:3Dzhao}.  The
  values chosen for the parameters $(a,b,c,r_0)$ for each galaxy are
  given in Table~\ref{tab:attributes}, together with the corresponding
  $\chi^2$ per degree of freedom. Our procedures for selecting these
  parameters are outlined in section ~\ref{sec:analysis}, where we
  also give references for the observational data plotted in each
  panel.}
\label{fig:Ir_cusp}
\end{figure}

In Figure~\ref{fig:Ir_cusp} we plot these surface density profiles for
each satellite on top of the observed data.  The scale radii vary over
the range $0.29~{\rm kpc~(Carina)} \leq r_0 \leq 0.67~{\rm kpc~(Fornax)}$ and
the outer slopes over the range $3.3~{\rm (Sextans)} \leq c \leq 7.5~{\rm
  (Leo)}$. As noted above, the degeneracies allow significant
variations in these quantities, particularly if $a$ and $b$ are
allowed to vary away from the values we have chosen. Our choices are
motivated in part by simplicity (e.g. for $a$), in part by
experimentation, determining which parameter ranges allow good fits
also to the kinematic data (see below).  For Sextans such
considerations lead us to settle on a relatively shallow outer density
profile, while for Leo the data force us to a steep outer
profile. Note that in all cases, the star counts were actually carried
out in elliptical annuli. The radial coordinate plotted is the
geometric mean of the major and minor axes which we expect to
correspond best to the count profile for circular annuli.  (The
typical ellipticities of these satellites are $\sim 0.3$
~\citep{Irwin:1995tb}).

\subsection{Kinematics} 
The kinematic datasets that we use consist of line-of-sight stellar
velocities from the samples of~\cite{Mateo:2007xh}
and~\cite{Walker:2008fc}. The latter use an ``Expectation
Maximization" method for evaluating membership and removing
contaminants from each sample, and we consider only those stars for
which~\cite{Walker:2008fc} assign $> 90\%$ probability of
membership. The resulting numbers of stars are listed in Table~
\ref{tab:attributes}. For Leo I, which is the only galaxy in our
sample without published membership probabilities, we use data from
~\cite{Mateo:2007xh}, and consider those stars as members that have
velocities in the range from 240 to 320~km~s$^{-1}$. As this range of
velocities is well separated from that of MW foreground stars, it is
unlikely that this sample suffers significant contamination. Other
methods for cleaning dSphs from contaminating MW halo stars have been
considered~\citep[e.g.][]{Klimentowski:2006qe}; these typically reduce
the velocity dispersion at outer radii. The \cite{Walker:2008fc}
membership cuts appear appropriate for our analysis here.

For each satellite, we bin the velocity data in a series of circular
annuli and estimate the mean square line-of-sight velocity in each
annulus as
\begin{equation} 
\hat \sigma^2  \equiv \langle v^2 \rangle - \langle e^2 \rangle.
\label{eq:hatsigma}
\end{equation}  
Here we define the velocity of a star as $v_\imath = v_{o,\imath} -
\overline{v}_o$, where $v_{o,\imath}$ is the observed velocity of the
$i^{th}$ star and $\overline{v}_o$ is the mean of these velocities over
all stars in the galaxy.  The quantity $e_\imath$ represents the
measurement uncertainty of the $i^{th}$ star, and angle brackets
represent an average over all the stars in a radial bin.  We further
assume that the error on $\overline{v}_o$ is negligible and that the actual
velocities are uncorrelated with their measurement error. With these
assumptions, $\hat\sigma^2$ is an unbiased estimator of the corresponding
population quantity, and approximating the sampling distributions of
$\langle v^2 \rangle$ and $\langle e^2 \rangle$ as normal,
the uncertainty on $\hat \sigma$ can be estimated as
\begin{equation}
\epsilon^2 = \frac{1}{2N} \frac{\langle v^2 \rangle^2}
{\langle v^2 \rangle - \langle e^2 \rangle}. 
\label{eq:epsilon} 
\end{equation} 

Given an estimate of the intrinsic velocity dispersion profile of each
satellite based on Eq.~\ref{eq:hatsigma}, we step through all the
subhalos in the six Aquarius simulations to determine which subhalo
has the (spherically averaged) potential that best describes the
data. Specifically, for each Aquarius subhalo, we derive a spherical
potential from the mass profile $M(r)$ and then use the Jeans
equation~(\ref{eq:jeansDE}) to calculate the line-of-sight velocity
dispersion profile, $\sigma_{los}(R)$, which corresponds to the model
star count profile of Table~\ref{tab:attributes} and an everywhere
isotropic velocity dispersion tensor. This line-of-sight velocity
dispersion is then averaged over the positions of all the stars in
each annulus to predict the population mean square velocity within that
annulus.  For each satellite-subhalo pair we then determine the
quantity
\begin{equation} 
\chi^2 = \sum_{\imath=1}^{N_{\rm bins}} \frac{[\hat \sigma_\imath-\sigma_{los}(R_\imath)]^2}{\epsilon_\imath^2}, 
\label{eq:chi2}
\end{equation} 
where $N_{\rm bins}$ is the number of annuli and $R_\imath$ is the
mean value of the projected radius of the stars in the $\imath^{th}$
annulus. For a given satellite, it then follows that the best fitting
Aquarius subhalo is the one that minimizes Eq.~\ref{eq:chi2}. 

Once a ``best'' subhalo has been identified in this way, we can
quantify whether it actually provides an acceptable fit by comparing
the $\chi^2$ value from Eq.~\ref{eq:chi2} to the theoretical
distribution of $\chi^2$ for $N_{\rm bins}$ degrees of freedom. If $p$
is the fraction of the theoretical distribution at larger values than
the measured $\chi^2$, then we can exclude the hypothesis that the
observed satellite has isotropic velocity dispersions and is hosted by
this ``best'' subhalo at confidence level $1-p$.  (Note that, given
our assumptions, there are no free parameters when comparing observed
and predicted dispersion profiles for a {\it specific} subhalo.) If
$p$ is not very small, then we conclude that the observed satellite
could be hosted by a $\Lambda$CDM subhalo. Note that the converse does
not apply. If $p$ is very small, the observed satellite could still
live in a $\Lambda$CDM subhalo if it has significant velocity
anisotropies.

\begin{figure}
\begin{center}
{\resizebox{7.5cm}{!}{\includegraphics{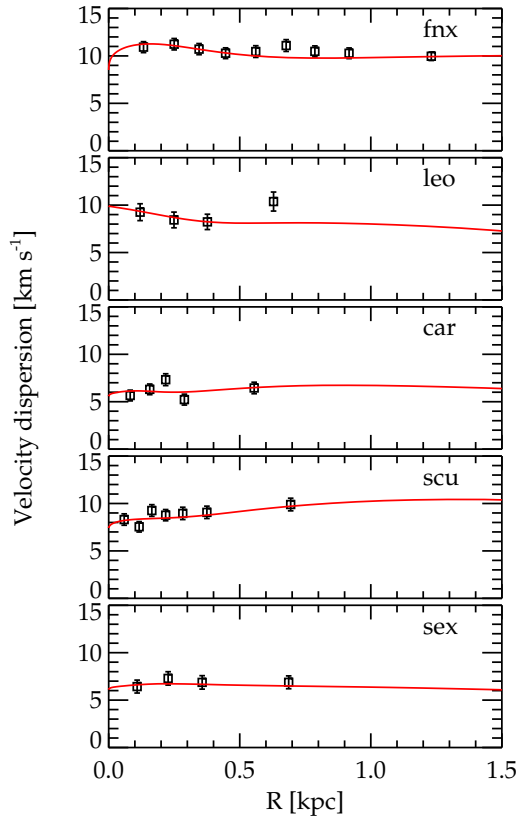}}} \\ 
\end{center}
\caption{Line-of-sight velocity dispersion for our five
satellites. The solid curves show the dispersion predicted by inserting the
  potential determined from the best fitting Aquarius subhalo and the
  photometric profile of Table~\ref{tab:attributes} into
  Eq.~\ref{eq:jeans}, assuming no velocity anisotropies.  The
symbols show the  observational data taken from ~\citet{Mateo:2007xh} (Leo I)
  and~\citet{Walker:2008fc} (Fornax, Carina, Sculptor, and Sextans).
  The errors on the velocity dispersion in each bin are assigned
  according to Eq.~\ref{eq:epsilon}. }
\label{fig:sigma}
\end{figure}

\begin{table}
\caption{\label{tab:attributes} Number of member stars with measured
  radial velocities in each of our five galaxies, together with the
  parameters in Eq.~\ref{eq:3Dzhao} for our preferred fits to their
  star count profiles, as shown in Fig.~\ref{fig:Ir_cusp}.  The final
  column gives the value of $\chi^2$ per degree of freedom for these
  count profile fits.  }
\begin{tabular}{lccccccccc}
\hline
Satellite & \# of stars&a&b&c&$r_0$ [kpc]&$\chi^2$/d.o.f\\
\hline
Fornax &2409&1&4&4.5&0.67&1.0\\
Leo I  &328&0&3&7.5&0.40&1.6\\
Carina &758&0.5&3&5.3&0.29&1.1\\
Sculptor &1392&0.5&3&5.5&0.32&0.4\\
Sextans &424&0.5&3&3.3&0.44&0.1\\
\hline
\end{tabular}
\end{table}

\section{Results} 
\label{sec:results}
In this section we turn to the implementation of the algorithms
described above.  We begin by finding the Aquarius subhalo that best
matches the line-of-sight velocity dispersion of each satellite under
the assumption of negligible velocity anisotropy and for the model
stellar density profile we have fitted to the observed counts. We then
check whether the line-of-sight velocity distributions of these models
are consistent with those observed.

\subsection{Best-fitting subhalos}
Figure~\ref{fig:sigma} compares the observed velocity dispersion
profiles of our five satellites to those predicted by
Eq.~\ref{eq:jeansDE} when a stellar system with a star count profile given
by Eq.~\ref{eq:3Dzhao} with the parameters in
Table~\ref{tab:attributes}, with a stellar mass-to-light ratio of 1, and with
negligible velocity anisotropy, is embedded in the Aquarius subhalo
that fits best according to the criterion of Eq.\ref{eq:chi2}.  The
p-values for these best-fitting subhalos are 0.6, 0.5, 0.6, 0.2 and
0.8 for Fornax, Leo I, Carina, Sculptor, and Sextans,
respectively. Thus in all cases the fit appears good from a
statistical point of view.

\begin{figure}
\begin{center}
{\resizebox{7.0cm}{!}{\includegraphics{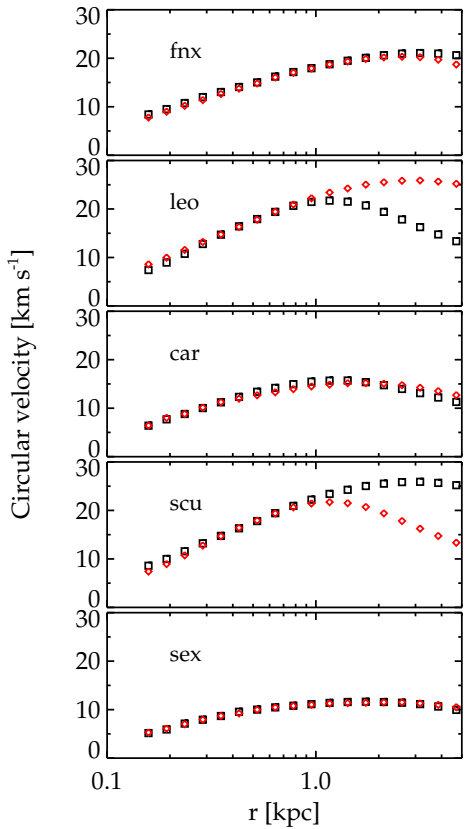}}} \\ 
\end{center}
\caption{Circular velocity profiles for the best-fitting subhalos
  (black squares) and the next-best-fitting subhalos (red diamonds)
  for our five satellites.  The symbols give 
  $\left[GM(r)/r)\right]^{1/2}$ which was used to calculate the
  potential from the dark matter component in the velocity dispersion
  profiles of Fig.~\ref{fig:sigma}.  The profiles of each pair of
  subhalos are almost identical over the range encompassed by the
  stellar kinematic data. At larger radii deviations can be large.  }
\label{fig:vc}
\end{figure}

It is important to note that when matching the data in
Fig.~\ref{fig:sigma}, the photometric parameters chosen in
Eq.~\ref{eq:3Dzhao} strongly affect the quality of the fit. For
example, for Fornax we are able to find an Aquarius subhalo that
matches both the photometry and the kinematics with an acceptable
p-value only if $a > 0.8$.  This motivates choosing $ a= 1$ for this
galaxy. For Sculptor, Carina, and Sextans the data constrain the
central slope of the stellar density profile less strongly, so for
these satellites we choose $a = 0.5$. As noted above, the
star count data for Leo I force a value of $a$ close to zero. Hence we pick
$a=0$ for simplicity, although a slightly negative $a$ gives a somewhat
better fit to the velocity dispersion data.

In Figure~\ref{fig:vc}, we show circular velocity curves for the two
``best'' subhalos for each of the satellites; the black squares
correspond to the subhalos plotted in Fig.~\ref{fig:sigma}. There is
essentially no difference between these pairs of curves at radii where
we have kinematic data.  This reflects the fact that, for our
assumptions, the potential can be derived directly from the
observational data so any ``acceptably fitting'' subhalo will have to
resemble the result of this exercise quite closely. At larger radii
the potential is effectively unconstrained, however, and the profiles
of the two subhalos can differ dramatically. For Fornax, Leo I,
Carina, Sculptor and Sextans the maximum circular velocities of the
best-fitting subhalos are 21, 22, 16, 26, 12 km~s$^{-1}$,
respectively.  The corresponding present-day dark matter masses are 7,
2, 2, 15, and $ 1 \times 10^8 \, M_\odot$. 

These results allow us to conclude that the Aquarius simulations
include at least one subhalo which is an acceptable host for
each of the five satellite galaxies, even under the restrictive
assumption of negligible velocity anisotropy.  We can extend this
analysis to estimate the total number of Aquarius subhalos that are
statistically consistent with the kinematic and photometric data for
each satellite. Specifically we count those subhalos that have a
p-value exceeding 10\%. (Remember we are still enforcing isotropic
velocity distributions and the specific profile parameters of
Table~\ref{tab:attributes}). For Fornax we find 13 acceptable
subhalos among the six simulations; for Sculptor there are 16 such
subhalos; for Carina there are 8 subhalos; for Leo~I there are 37; 
and for Sextans we find over 100 acceptable
hosts. For Sextans this large number 
of hosts is a reflection both of the fact that its velocity
dispersion is lower and that the errors on the data are relatively
large.

\begin{figure*}
\begin{center}
\begin{tabular}{ccc}
{\resizebox{6.1cm}{!}{\includegraphics{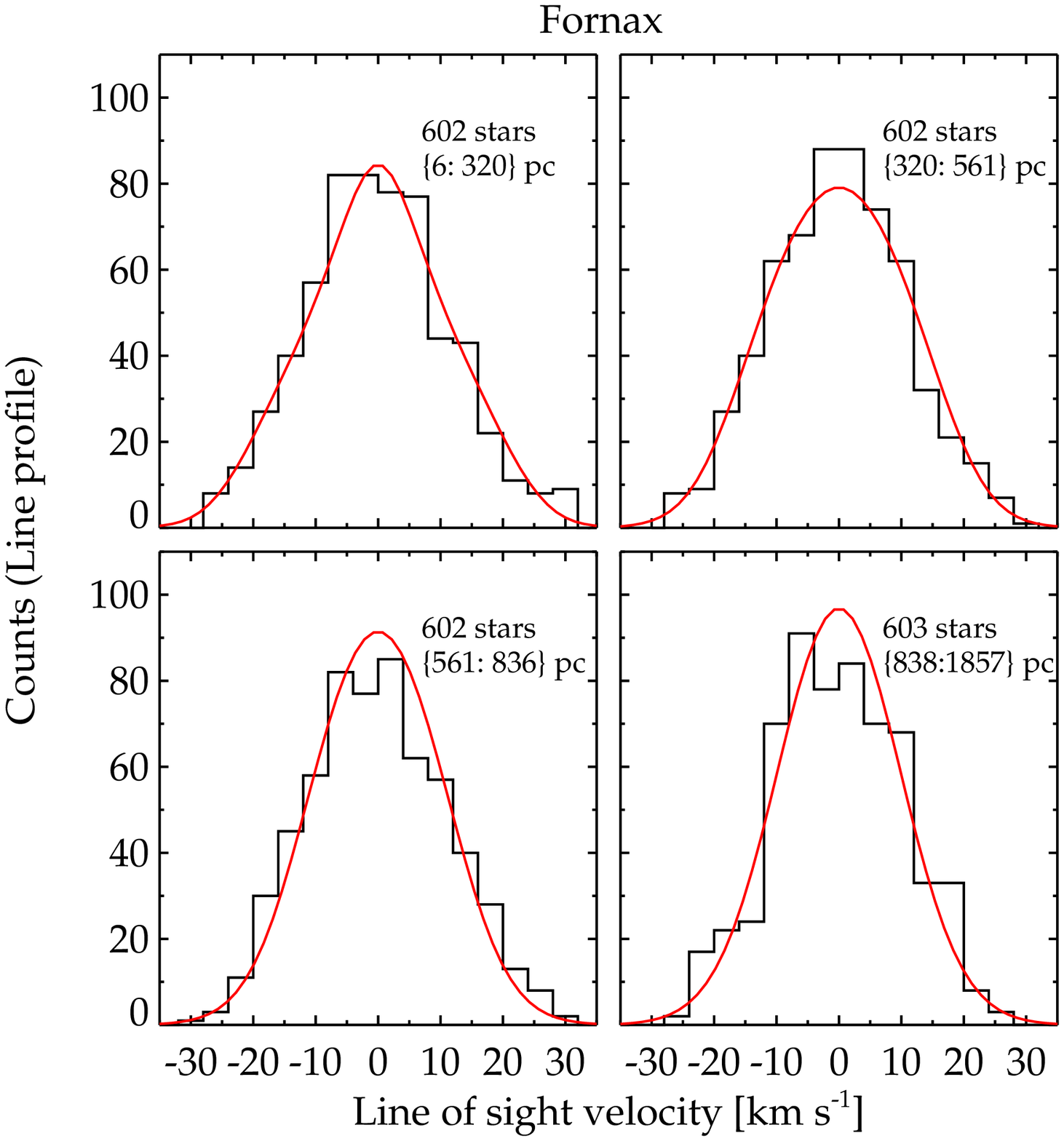}}} &
 {\resizebox{6.1cm}{!}{\includegraphics{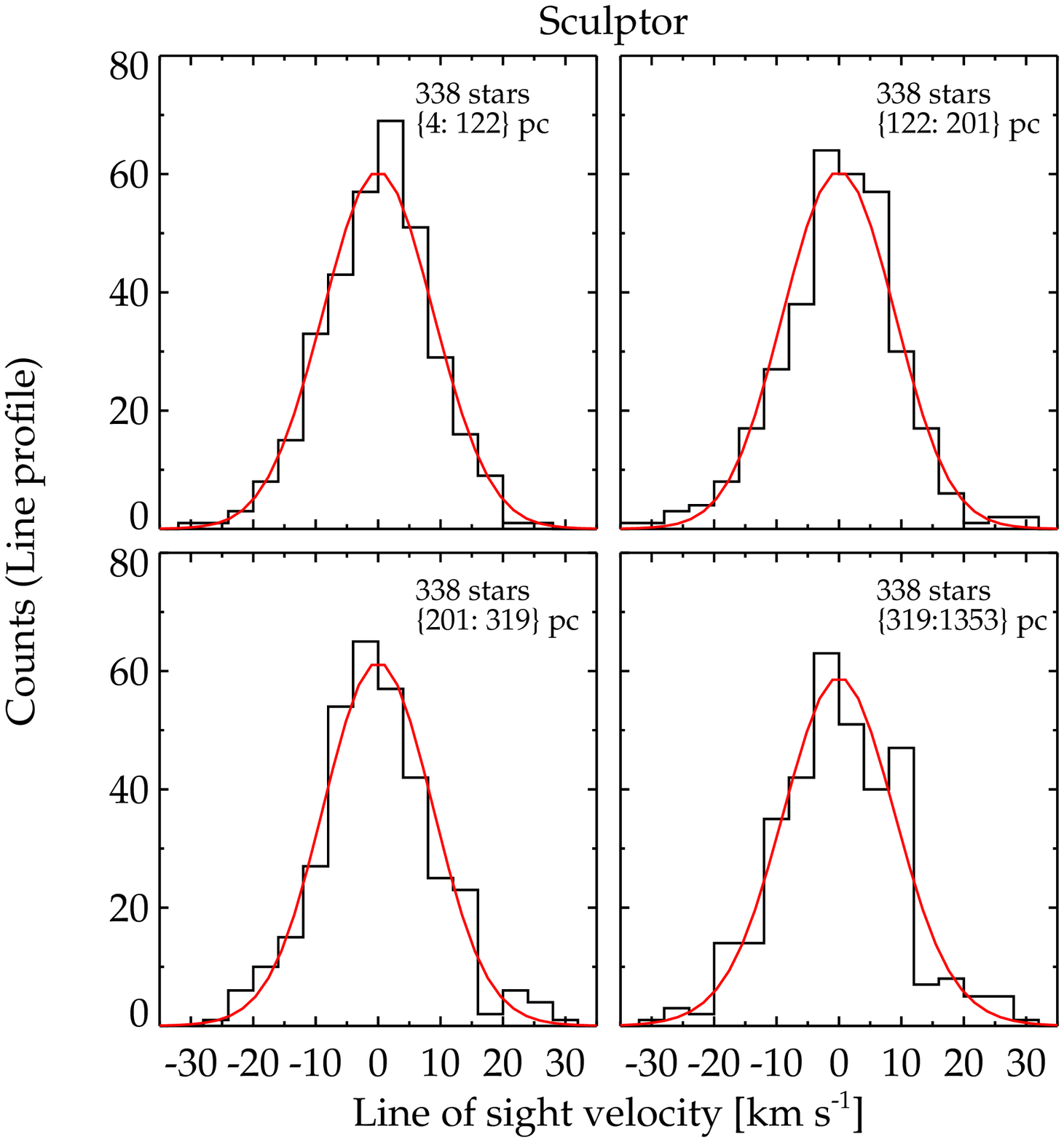}}} & 
{\resizebox{6.1cm}{!}{\includegraphics{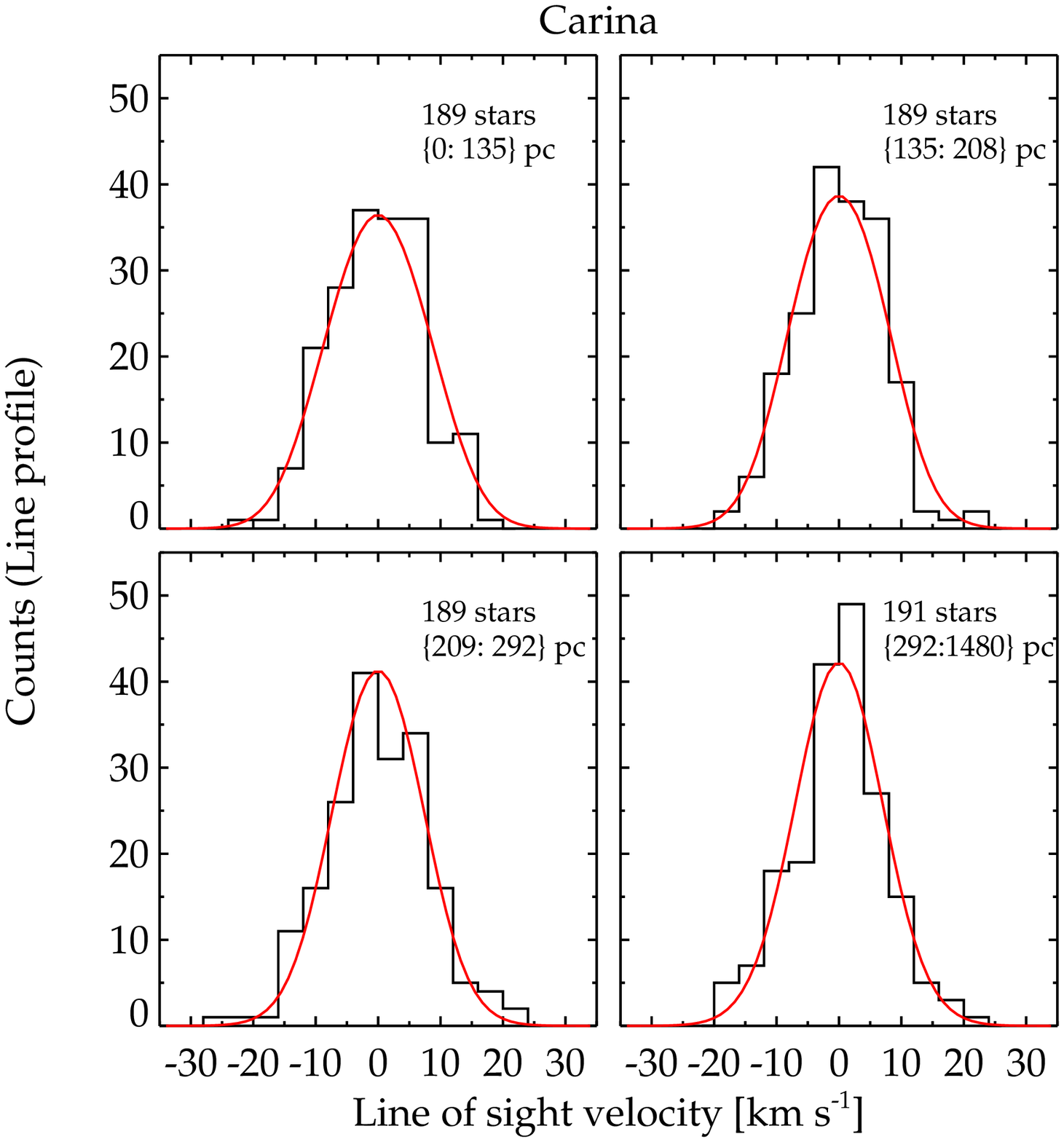}}} \\ 
\end{tabular}
\begin{tabular}{cc}
{\resizebox{6.1cm}{!}{\includegraphics{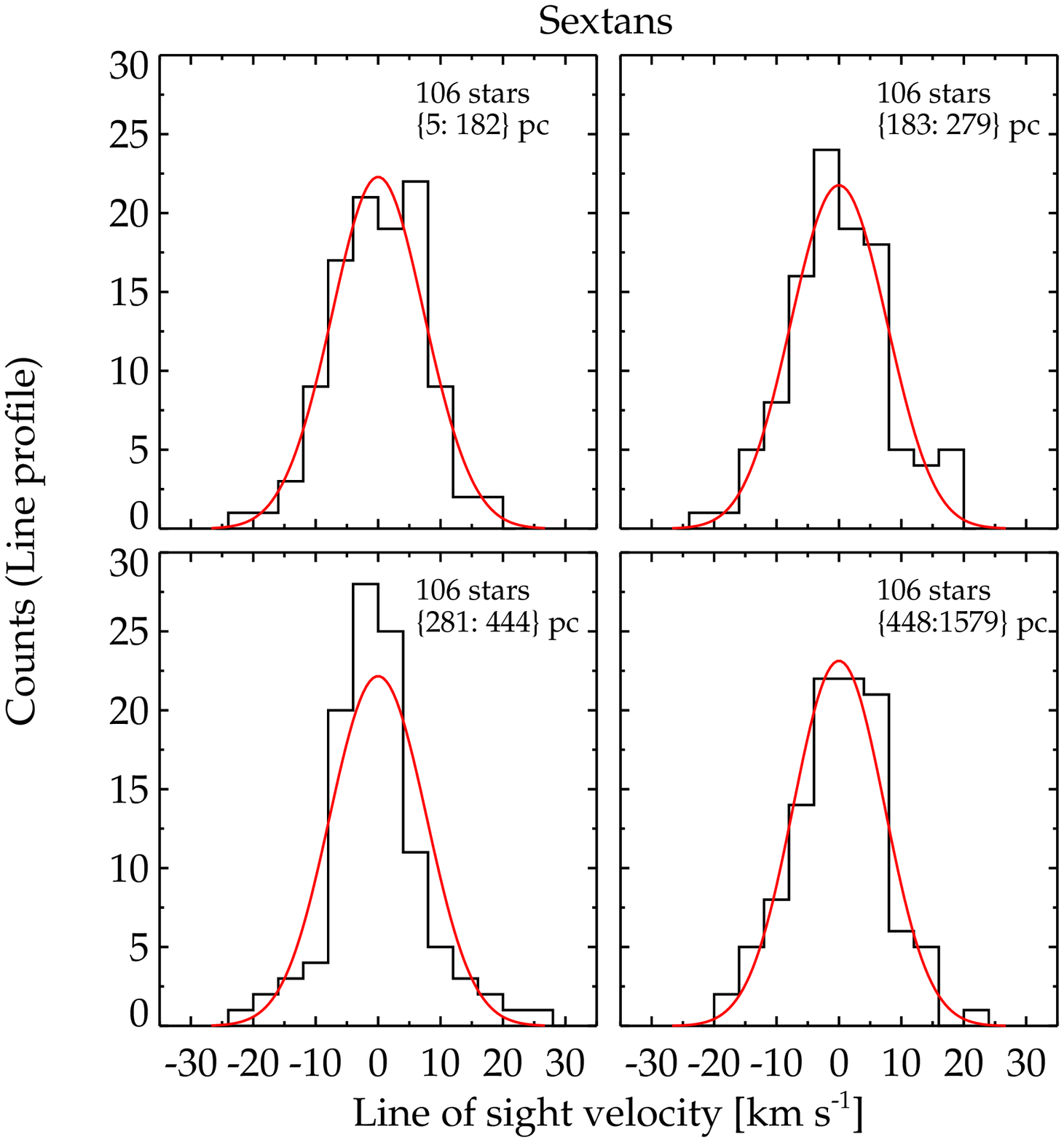}}} & 
{\resizebox{6.1cm}{!}{\includegraphics{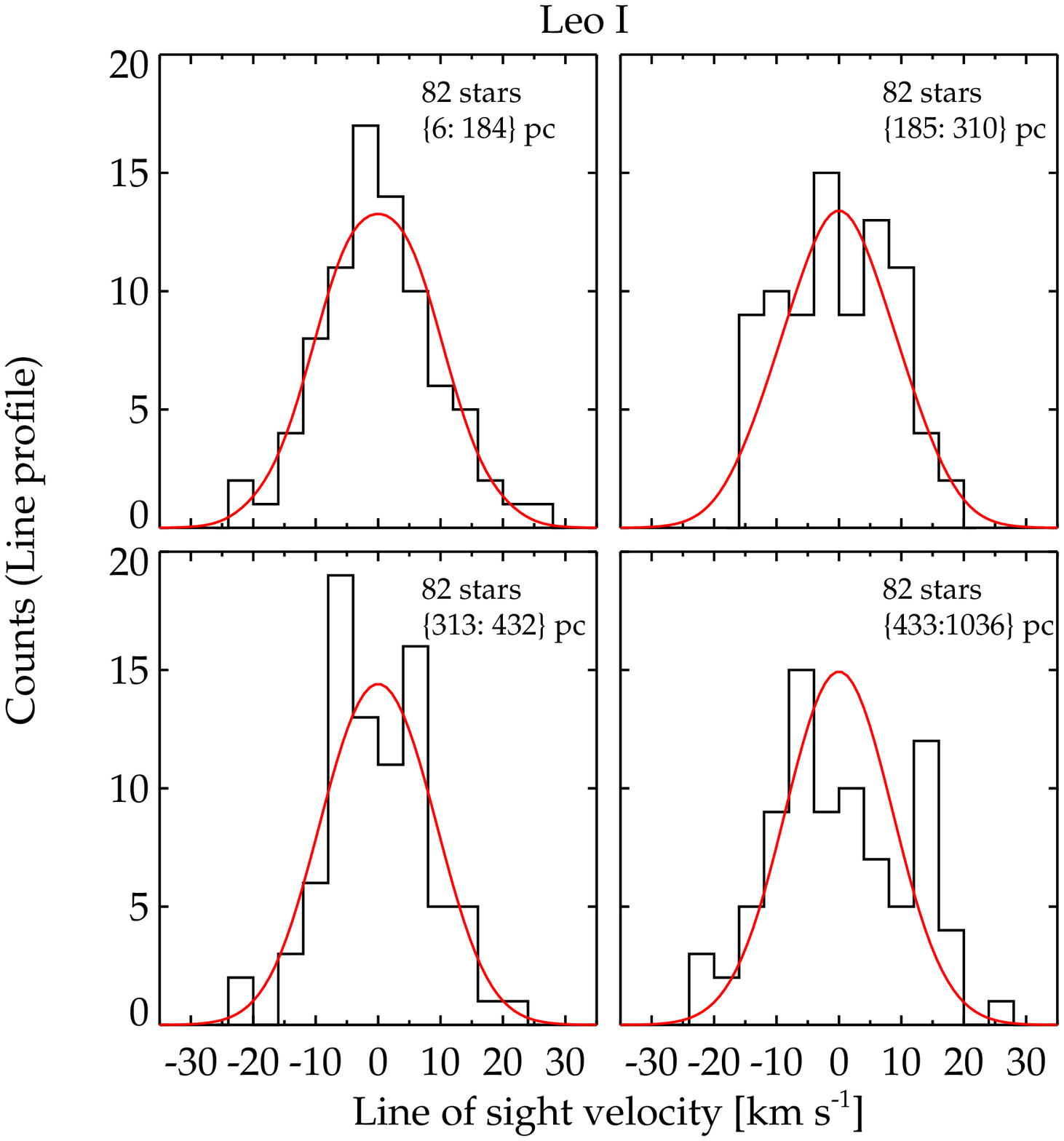}}} \\
\end{tabular}
\end{center}
\caption{Velocity distributions in four bins for each of our
  satellites.  In each panel, a solid curve shows the theoretical
  distribution averaged over the radial positions of all the stars in
  the bin, and then smoothed with a Gaussian representing the typical
  observational error on the stellar velocities.  Labels at the upper
  right list both the total number of stars in each bin and the
  approximate radial range they encompass.  }
\label{fig:hists_r_4bins}
\end{figure*}

\subsection{Full velocity distributions} 
The preceding analysis shows that $\Lambda$CDM subhalo potentials can fit
the star count and line-of-sight velocity dispersion profiles of Milky
Way satellites to good accuracy, even under the assumption of
negligible velocity anisotropy. We now use the techniques described in
Section~\ref{sec:modeling} to study if these same models provide good
descriptions of the full line-of-sight velocity distributions, for
example, of their higher order moments.

The line-of-sight velocity distribution at projected radius $R$ is a
convolution of the intrinsic distribution from Eq.~\ref{eq:fhat}
with the measurement uncertainty of the stellar velocities. The former
has to be averaged over the radial positions of all the stars in each
of our annuli, and we model the latter by a Gaussian with dispersion
equal to the mean quoted velocity error of these same stars. This
measurement uncertainty ranges from $\sim 3$ km s$^{-1}$ for all the
Carina bins and the innermost Sculptor bins, down to $\sim 1.5$ km
s$^{-1}$ for all the Fornax bins.

The resulting line-of-sight velocity distributions are shown for each
of our five galaxies in Fig.~\ref{fig:hists_r_4bins}.  For each galaxy
we used four circular annuli containing nearly equal numbers of
stars. Our motivation for this binning scheme is to retain some
information about the variation of distribution shape with radius
while keeping a large number of stars in each bin in order to better
constrain the shape of the distribution. This results in fewer bins in
Fig.~\ref{fig:hists_r_4bins} than in Fig.~\ref{fig:sigma} for Fornax,
Sculptor, and Carina.  In each panel of Fig.~\ref{fig:hists_r_4bins}
the solid curve is the line-of-sight velocity distribution calculated
as above and normalized to have the same area as the corresponding
histogram.

To quantify the level of agreement between model and data in
Fig.~\ref{fig:hists_r_4bins}, we compare the distributions of
$|v_\imath|$ in each panel using a KS test. We take the modulus here
because the distribution of the line-of-sight velocity relative to the
galaxy mean is expected to be symmetric about zero for {\it any}
equilibrium model (even rotating and/or non-spherical ones) after
averaging over a circular annulus.  As a result, all shape information
is contained in the distribution of $|v_\imath|$, and restricting
the test in this way enhances its sensitivity to higher order moments.
The maximum difference between the normalized cumulative distributions
of $|v_\imath|$ for data and model is then a measure of the
confidence level at which we can reject the null hypothesis that
our simple isotropic, spherical model represents the full, observed
line-of-sight velocity distribution in the annulus.

Results of this KS test for each of the four annuli and for all of our
satellites are shown in Table~\ref{tab:ks}, with bins 1-4 ordered by
increasing radius. These values indicate that our predicted velocity
distributions are generally in good agreement with the data.  The
annuli with the lowest probabilities are bins 3 and 4 of Fornax and
bin 4 of Leo I; for Fornax bin 3 and Leo I bin 4 the null hypothesis
can be excluded with $>99\%$ confidence. As can be seen in
Fig.~\ref{fig:hists_r_4bins}, the measured line profiles appear less
peaked (platykurtic) than the models in these annuli. For Fornax the
effect is quite weak, but is nevertheless significantly detected
because of the large number of stars involved. For all the other
panels of Fig.~\ref{fig:hists_r_4bins} differences are less
significant and are quite small. Note that the kurtosis of these
distributions is expected to be quite sensitive to velocity
anisotropy, so the fact that our models fit fairly well can be taken
as an indication that anisotropies are probably weak.

A more direct measurement of the kurtosis of the line-of-sight
velocity distributions can be obtained by estimating their
fourth moment directly. For each annulus, we calculate a sample
kurtosis from the $N$ stars it contains as 
\begin{equation} 
\kappa = {\langle v^4 \rangle} /{\langle v^2 \rangle^2} -3. 
\label{eq:kurtosis} 
\end{equation} 
The kurtosis is defined so that a Gaussian model gives $\kappa=0$.  We
approximate the uncertainty of the sample kurtosis by $\sqrt{24/N}$, the
scatter expected for random samples from a normal distribution. For
our theoretical model, we calculate second and fourth moments
from Eq.~\ref{eq:moment}, after smoothing to account for measurement
errors, and we then substitute these into Eq.~\ref{eq:kurtosis} to
obtain the predicted kurtosis.

\begin{figure}
\begin{center}
{\resizebox{7.0cm}{!}{\includegraphics{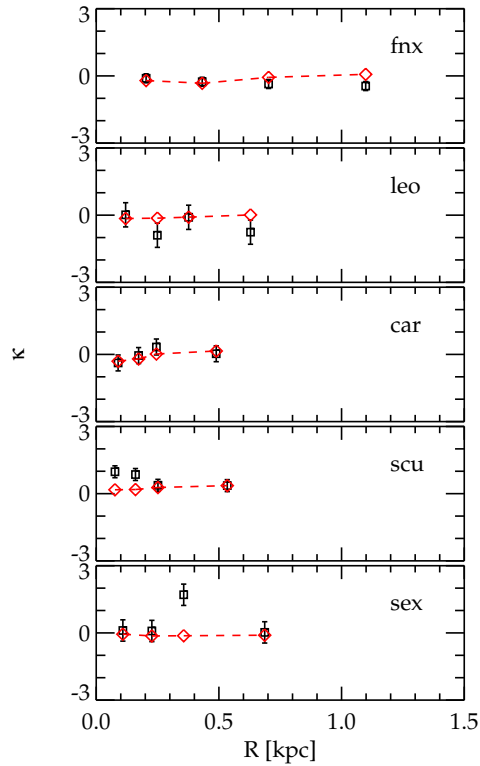}}} \\ 
\end{center}
\caption{Comparison of the predicted and observed kurtosis of the
  line-of-sight velocity distribution in four equally populated annuli
  in each of our five satellites. The kurtosis is defined to be zero
  for a normal distribution.  The black squares are direct
  observational estimates for each annulus, with errors given by
  $\sqrt{24/N}$, where $N$ is the number of stars.  The red diamonds
  are theoretical predictions from our isotropic models after
  convolution with the measurement errors and averaging over the
  stellar positions in each annulus.  }
\label{fig:kurtosis}
\end{figure}

The results of this exercise are shown in
Fig.~\ref{fig:kurtosis}. The models predict very little kurtosis in
almost all annuli, and the data agree with this in most cases.
Comparing Fig.~\ref{fig:kurtosis} to the second moment data in
Fig.~\ref{fig:sigma} clarifies the origin of the discrepancies
uncovered by our KS test. For example, in the outer annuli of
Fornax, the model distribution is (slightly) narrower than the data
while the kurtosis estimates are almost in agreement. For the 4th
annulus of Leo I, the model RMS is again smaller than for the data,
and the data are more platykurtic. For Leo I, the 2nd annulus also
shows a marginally significant platykurtic signal, while for Sextans
bin 3 the data are more leptokurtic than the model.  However, in the
latter two instances the number of stars is too small for the KS
test to indicate a significant discrepancy. The rather large
bin-to-bin fluctuations in the observational estimates of kurtosis
suggest that our Gaussian error bars may be underestimating the true
sampling uncertainties.

From this analysis, we conclude that the observed line-of-sight
velocity distributions agree surprisingly well with our spherical,
isotropic models embedded in $\Lambda$CDM subhalo potentials. The
remaining differences can plausibly be ascribed to departures from
isotropy, from spherical symmetry, from dynamical equilibrium, or
(more likely) from a combination of these. Detailed modeling is, of
course, necessary to test this possibility, but it is beyond the scope
of the present paper. Our main goal here has been to show that
simulated $\Lambda$CDM halos contain subhalos with potentials
consistent with those hosting the observed stellar populations of
Milky Way satellites, even if these are assumed to have negligible
velocity anisotropy.

\begin{table}
\centering
\caption{\label{tab:ks} KS probabilities for the maximum difference
  between the observed and modeled cumulative distributions of $|
  v_\imath|$ within four equally populated annuli in each of our
  observed satellites.  Bins 1-4 correspond to the annuli of
  Fig.~\ref{fig:hists_r_4bins} ordered from inside to outside.  The
  complement of each of these values represents the confidence level
  at which the hypothesis that the data are drawn from the theoretical
  distribution can be rejected.}
\begin{tabular}{lllll}
\hline
Satellite & bin 1 & bin 2& bin 3& bin 4\\
\hline
Fornax &0.05&0.15&$0.003$&0.02\\
Leo I  &0.46&0.52&0.31&0.003\\
Carina &0.51&0.44&0.29&0.34\\
Sculptor &0.32&0.68&0.62&0.67\\
Sextans &0.41&0.81&0.03&0.97\\
\hline
\end{tabular}
\end{table}

\section{Discussion and Conclusion} 
We have investigated whether the gravitational potentials of subhalos
in N-body simulations of $\Lambda$CDM halo formation are consistent
with the high-quality photometric and kinematic data available for
five of the brighter satellites of the Milky Way. We find that a
direct mapping is, in fact, possible between each of these 
satellites and a subset of the dark matter subhalos in the six high
resolution simulations of the Aquarius Project.  Star count profiles
with inner cusps scaling as $r^{-a}$ with $0\leq a \leq 1$ can provide
good fits to the observed counts. Placed in the measured Einasto-like
potentials of appropriately selected subhalos, they also fit the
observed, nearly flat line-of-sight velocity dispersion profiles very
well, even under the restrictive assumption of negligible velocity
anisotropy.  Such isotropic models fit the {\it shapes} of the
observed line-of-sight velocity distributions well, in addition to
their second moments.

We have measured the present-day maximum circular velocities of the
``best-fit'' subhalos for each of these five satellites. These range
from 10 to 30~km~s$^{-1}$. Subhalos consistent with hosting the
observed systems at $z=0$ have peak circular velocities (i.e. the
largest maximum circular velocity they {\it ever} had) ranging from 12
to 50~km~s$^{-1}$.  The maximum past masses of the main progenitors of
these subhalos range up to $\sim 5\times 10^9$ M$_\odot$, while their
masses at $z=7$ (the approximate lower bound on the redshift of
reionization~\citep{Dunkley:2008ie}) range up to $\sim 10^9$
M$_\odot$. At $z=0$ their Galactocentric distances range from 40 to
400~kpc.

Our results indicate that current data on faint Milky Way satellites
are consistent with these galaxies living in $\Lambda$CDM halos. They
do not, however, explain {\it why} galaxies living in such subhalos
should have the observed properties. Exploring this issue is an  
important task for future work
\citep[e.g.][]{Li:2009kv,Cooper:2009kx,Sawala:2009,Okamoto:2009rw,Busha2010}.

\section*{Acknowledgments} 
We thank Matt Walker for helpful discussions and for making the
velocity data for the satellites publically available. We thank
V. Smolcic and G. Battaglia for providing Leo I and Sculptor
photometry data respectively. We also thank Jie Wang for assistance
with the Aquarius simulations data. The Aquarius project is part of
the programme of the Virgo Consortium for cosmological
simulations. Support for LS for this work was provided by NASA
through Hubble Fellowship grant HF-51248.01-A awarded by the Space
Telescope Science Institute, which is operated by the Association of
Universities for Research in Astronomy, Inc., for NASA, under
contract NAS 5-26555. CSF acknowledges a Royal Society Wolfson
Research Merit Award. This work was also supported in part by an
STFC rolling grant to the ICC.

\bibliography{kinematics} 

\end{document}